\documentclass[aps,pra,twocolumn,amsmath,amssymb,nofootinbib,superscriptaddress]{revtex4-2}
\usepackage[colorlinks,
           linkcolor=blue,
            anchorcolor=blue,
            allcolors = blue,
            citecolor=blue
            ]{hyperref}

\usepackage{bm}
\usepackage{graphicx}
\usepackage{enumerate}
\usepackage{enumitem}
\usepackage{lipsum}

\begin{document}

% \title{Scalable Superconducting Processor with $99.9\%$ CZ Gate Fidelity via High-Precision Calibration}
% \title{Achieving 99.9\% CZ Gate Fidelity on Scalable Superconducting Processors via an AI-Friendly Calibration Workflow}
\title{High-Precision Calibration Workflow Achieves Above $99.9\%$ CZ Gate Fidelity on a Scalable Superconducting Processor}

% \title{Achieving 99.9\% Fidelity on Scalable Superconducting Processors via Calibration Workflow }

\author{Huili Zhang}\thanks{These authors contributed equally to this work.}
\affiliation{Beijing Key Laboratory of Fault-Tolerant Quantum Computing, Beijing Academy of Quantum Information Sciences, Beijing 100193, China}

\author{Meiling Li}\thanks{These authors contributed equally to this work.}
\affiliation{Beijing Key Laboratory of Fault-Tolerant Quantum Computing, Beijing Academy of Quantum Information Sciences, Beijing 100193, China}

\author{Shuang Yang}
\affiliation{Beijing Key Laboratory of Fault-Tolerant Quantum Computing, Beijing Academy of Quantum Information Sciences, Beijing 100193, China}

\author{Yaqing Feng}
\affiliation{Beijing Key Laboratory of Fault-Tolerant Quantum Computing, Beijing Academy of Quantum Information Sciences, Beijing 100193, China}

\author{Yulong Li}
\affiliation{Beijing Key Laboratory of Fault-Tolerant Quantum Computing, Beijing Academy of Quantum Information Sciences, Beijing 100193, China}

\author{Cheng Chen}
\affiliation{Beijing Key Laboratory of Fault-Tolerant Quantum Computing, Beijing Academy of Quantum Information Sciences, Beijing 100193, China}

\author{Pei Liu}
\email{liupei@baqis.ac.cn}
\affiliation{Beijing Key Laboratory of Fault-Tolerant Quantum Computing, Beijing Academy of Quantum Information Sciences, Beijing 100193, China}

\author{Guangming Xue}
\affiliation{Beijing Key Laboratory of Fault-Tolerant Quantum Computing, Beijing Academy of Quantum Information Sciences, Beijing 100193, China}
\affiliation{Hefei National Laboratory, Hefei 230088, China}

\author{Haifeng Yu}
\email{hfyu@baqis.ac.cn}
\affiliation{Beijing Key Laboratory of Fault-Tolerant Quantum Computing, Beijing Academy of Quantum Information Sciences, Beijing 100193, China}
\affiliation{Hefei National Laboratory, Hefei 230088, China}

\date{\today} 

\begin{abstract}
High-fidelity universal two-qubit gates are critical for building fault-tolerant quantum computers. In scalable superconducting processors, shortened coherence times introduce more incoherent errors in gate operations.
With a constrained error budget, there is reduced tolerance for coherent errors stemming from parameter deviations.
In this work, we develop a closed-loop workflow to enhance the CZ gate calibration precision.
Utilizing the echoed leakage error amplification (ELEA) and the repurposed context-aware fidelity estimation (CAFE) circuits, we suppress the population leakage
to non-computational states, and, for the first time, demonstrate a CZ gate fidelity exceeding $99.9\%$ on an 84-qubit processor, with coherent error suppressed to 0.007\%.
Meanwhile, we obtain a median fidelity of $99.25\%$ among 72 CZ gates, demonstrating that the workflow can be generalized to the calibration of parallel CZ gates.
Finally, we realize automated calibration and observe enhanced stability of the CZ gate throughout 9-hour comparative monitoring experiments. Our results, realized on a completely domestic platform, establish an efficient and automated route to quantum computation with superconducting quantum systems. 
\end{abstract}

\maketitle

\section{Introduction}

Over the past decade, superconducting quantum systems have been widely applied in quantum simulation \cite{aspuru2005simulated, jin2025topological, xu2024non, bauer2023quantum, google2020hartree, jiang2026one}, metrology \cite{zhang2026distributed, niu2023low, 2023gusary, beaulieu2025criticality, Danilin_2024}, and quantum machine learning \cite{zhang2026experimental, hu2019quantum, Hall2024, genois2025quantum, convy2022machine}.
To achieve the ultimate goal of fault-tolerant quantum computation, a practical approach is to embed logical qubits in physical qubits, i.e., quantum error correction (QEC) \cite{Shor1995, Steane1997, bravyi1998quantum, dennis2002topological, breuckmann2021quantum, wang2026demonstration}.
The number of qubits and fidelity of gate operations are two of the most important figures of merit for realizing QEC.
It is believed that QEC becomes feasible when the error rates of physical qubits are around $0.1\%$ \cite{gidney2025factor, evered2026high}.
In the field of superconducting qubits, with advances in fabrication, chip design, and pulse optimization, CZ gates around this error rate have been achieved on both fluxonium qubits \cite{ding2023high, lin2025stable, zhang2024tunable} and transmon qubits \cite{li2024realization, marxer2025above}.

Nevertheless, these fidelities have been demonstrated on qubits fabricated on isolated superconducting samples.
Achieving comparable CZ gate fidelities on scalable superconducting processors remains non-trivial.
Gate errors, based on the noise source, can be categorized into incoherent errors and coherent errors \cite{gross2024characterizing, wei2024characterizing}.
Incoherent errors result from energy relaxation and dephasing processes in qubits, while coherent errors originate from imperfect parameters during calibration.
As the fabricated surface area increases with chip size, qubits become more susceptible to two-level system (TLS) defects at metal-substrate and metal-oxide interfaces, introducing additional energy relaxation channels \cite{crowley2023disentangling, muller2019towards, woods2019determining, Wang_2022}.
Meanwhile, the higher density of control lines creates additional depolarizing channels \cite{kono2020breaking, krinner2019engineering}. 
These problems inevitably introduce more incoherent errors into the CZ gate, reducing the error budget for coherent errors. 

In this vein, we develop a closed-loop calibration workflow for suppressing the coherent component of the CZ gate error in this work. 
The standard error amplification (SEA), the echoed leakage error amplification (ELEA), and the repurposed context-aware fidelity estimation (CAFE) \cite{debroy2023context} circuits are applied in alternation in each step of the workflow. We validate this workflow on the superconducting processor \href{https://quafu-sqc.baqis.ac.cn/home}{$Shenglian$} \cite{baqisQuafuSuperconducting}, which is composed of 84 tunable transmon qubits coupled via tunable transmon couplers. After calibration, we obtain a median fidelity of $(99.25\pm0.01)\%$ among 72 CZ gates. In particular, the maximum CZ gate fidelity is $(99.92\pm0.01)\%$, with $0.007\%$ attributable to coherent errors. 
Finally, we perform the calibration automatically, thereby enhancing the stability of the CZ gate in a 9-hour monitoring experiment.

\begin{figure*}[t]
    \centering
    \includegraphics[]{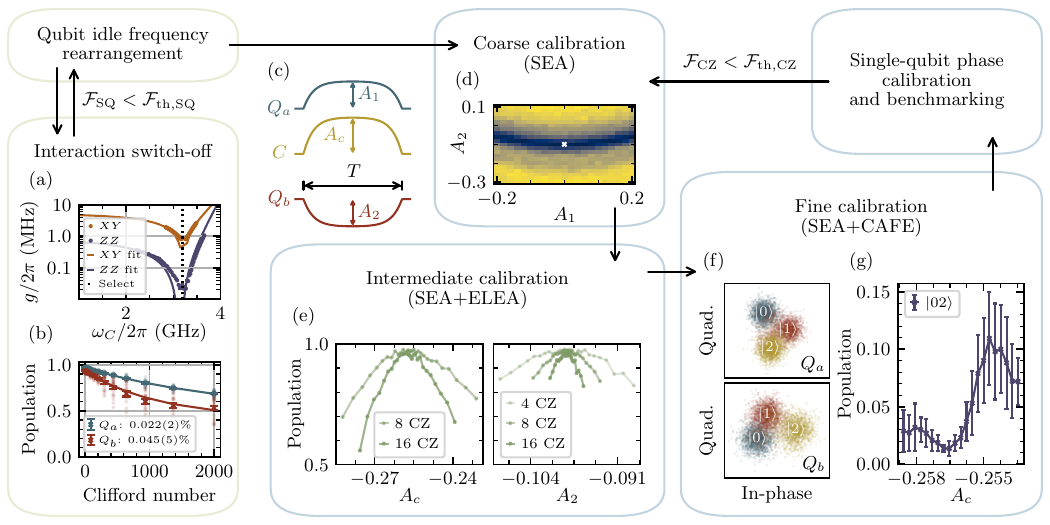}
    \caption{The workflow and results for calibrating CZ gates. (a) The experimental results of the effective $XY$ interaction (red) and $ZZ$ interaction (blue) between $Q_{a}$ and $Q_{b}$ versus coupler frequency $\omega_C$.
    The coupler idle frequency is biased at $\omega_{c}/2\pi\approx3.2$~GHz (dashed line) to reduce the interaction strength.
    (b) Results for characterizing single-qubit gates with randomized benchmarking (RB).
    The fitted errors per Clifford gate are $(0.022\pm0.002)\%$ and $(0.045\pm0.005)\%$ for $Q_{a}$ (blue) and $Q_{b}$ (red), respectively.
    (c) The flux pulses to implement the CZ gate with duration $T$. The flux pulse amplitudes of $Q_1$, $Q_2$ and $C$ are labeled as $A_1$, $A_2$, and $A_c$, respectively.
    (d) Population of qubit $Q_{a}$ as a function of $A_1$ and $A_2$.
    The dark region indicates a conditional phase on the state $|11\rangle$.
    (e) Progressive tuning of the pulse amplitude $A_c$ and $A_2$ with an increasing CZ number $2n$. The standard error amplification (SEA) and echoed leakage error amplification (ELEA) circuits are applied iteratively. 
    (f) Three-state single-shot readout clouds for $Q_{a}$ and $Q_{b}$. (h) Population of state $|02\rangle$ as a function of $A_c$ with 60 CZ gates in the context-aware fidelity estimation (CAFE) circuit.
    Error bars represent the standard error of CAFE circuits with 10 different initial state.}
    \label{Fig1}
\end{figure*}

\section{Calibration workflow}

We start with a minimal system that can produce entanglement, consisting of nearest-neighbor qubits $Q_{a}$, $Q_{b}$, and coupler $C$. Initially, the timing order among the qubit microwave drive, qubit flux, and coupler flux is calibrated, and flux distortions are compensated via infinite impulse response filters. We subsequently characterize the coherence properties versus the qubit frequency. The qubit idle frequencies are arranged by considering energy relations, dephasing and operation imperfections, and are modified when the single-qubit gate fidelity is less than a threshold $\mathcal{F}_{\text{th,SQ}}$. Here, the idle frequencies of $Q_a$ and $Q_b$ are $\omega_{{a}}/2\pi\approx 4.224$~GHz and $\omega_{{b}}/2\pi\approx4.411$~GHz. The effective interaction between $Q_{a}$ and $Q_{b}$ manifests as $XY$ interaction and $ZZ$ interaction, both can be tuned by adjusting the coupler frequency $\omega_{C}$ \cite{ding2025pulse, yan2018tunable}.
We bias the coupler at $\omega_{C}/2\pi\approx3.2$~GHz, where both interaction strengths are minimized, as shown in Fig.~\ref{Fig1}~(a).
The interaction strengths of next-nearest-neighbor qubits are estimated to be less than $2\pi\times0.1$~MHz, thus are neglected in the calibration.

For single-qubit gates, we employ the $U_3$ decomposition with the need for calibrating $X/2$ gate. We implement the single-qubit gate via a microwave pulse with a raised-cosine envelope and a duration of $14$~ns.
We use the derivative removal by the adiabatic gate (DRAG) technique to suppress leakage and phase errors \cite{Motzoi_2009, Chen_2016}.
The randomized benchmarking (RB) results are shown in Fig.~\ref{Fig1}~(b), with fitted errors per Clifford gate of $(0.022\pm 0.002)\%$ and $(0.045\pm0.005)\%$ for $Q_{a}$ and $Q_{b}$, respectively. 

% The single-qubit gate fidelity is above the setting threshold $\mathcal{F}_{\text{th,SQ}}$, 

\begin{figure}[t]
    \centering
    \includegraphics[]{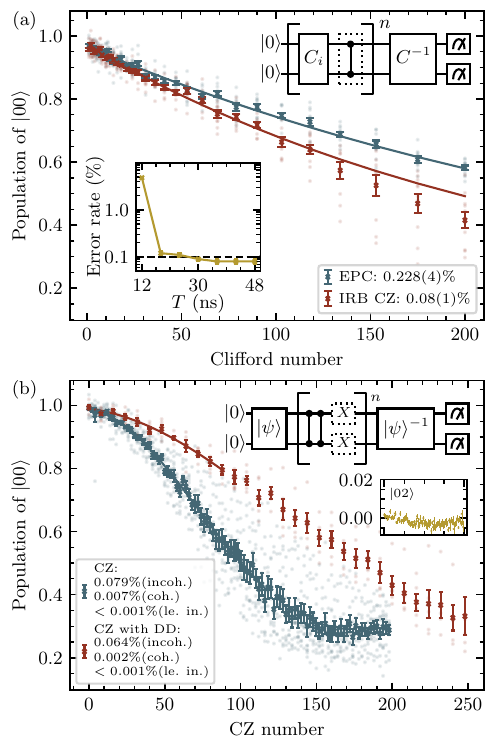}
    \caption{Error analysis of CZ gate after calibration. (a) Standard randomized benchmarking (SRB) and interleaved randomized benchmarking (IRB) results of CZ gate on $Q_{a}$ and $Q_{b}$, with the sequences shown in the inset. The calculated error per Clifford (EPC) of $(0.228\pm0.004)\%$ in SRB. 
    The CZ gate fidelity, derived from a comparison between SRB and IRB, is $(99.92\pm0.01)\%$. The CZ gate error rates at each gate duration $T$ are shown in another inset. (b) Population of state $|00\rangle$ as a function of CZ gate number, using the CAFE and CAFE with dynamical decoupling (DCAFE) circuits (inset). The population of state $|02\rangle$ versus CZ gate number is shown in another inset. The fitting results are 99.914\% and 99.334\% for CZ gates in CAFE and DCAFE circuits, respectively. Error bars represent the standard error of 10 CAFE circuits with different initial states $|\psi\rangle$.}
    \label{Fig2}
\end{figure}

\begin{figure*}[t]
    \centering
    \includegraphics[]{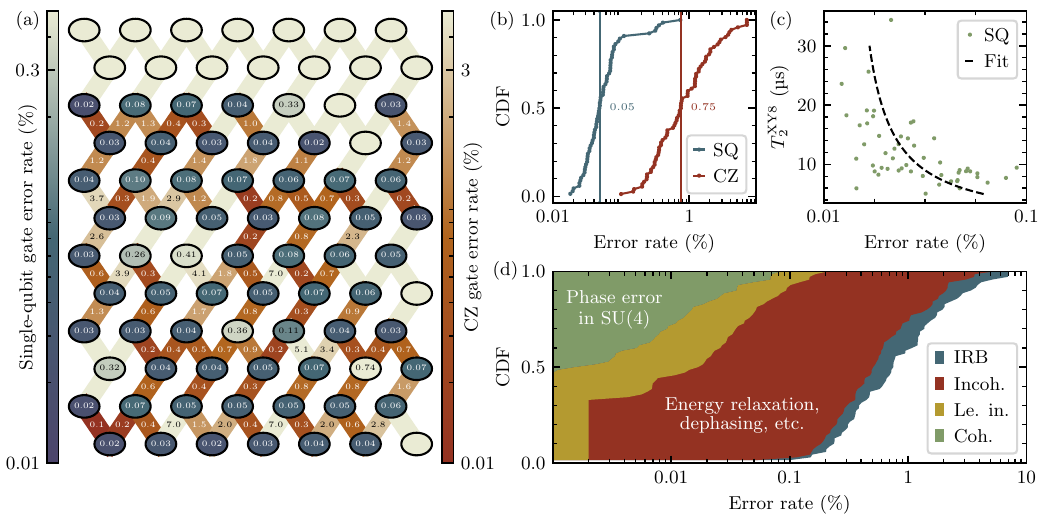}
    \caption{Error analysis of parallel CZ gates on a scalable superconducting processor.
    (a) Heatmap of parallel single-qubit (SQ) error rates and CZ gate error rates on the 84-qubit quantum processor.
    (b) Cumulative distribution function (CDF) of SQ gate and CZ gate error rates.
    Two vertical lines indicate the median error rates of $0.05\%$ and $0.75\%$, respectively.
    (c) The error rate of SQ gates versus $T_{2}^{\rm XY8}$, fitted by an exponential decay with the inverse dephasing time.
    (d) CDF of the average error rates measured from $72$ CZ gates.
    The overall error calculated by IRB (blue) encompasses incoherent errors stemming from energy relaxation and dephasing (red), incoherent errors induced by leakage (orange), and coherent errors arising from phase errors in SU(4)(green).}
    \label{Fig3}
\end{figure*}

The CZ gate is realized via the non-adiabatic scheme. In this scheme, qubit frequencies are tuned to bring the states $|11\rangle$ and $|02\rangle$ into resonance, and the coupler frequency is tuned to enhance the $XY$ interaction strength between states $|11\rangle$ and $|02\rangle$ simultaneously.
The full oscillation between the states $|11\rangle$ and $|02\rangle$ yields a relative phase $\pi$ on the state $|11\rangle$~\cite{dicarlo2010preparation, strauch2003quantum, yamamoto2010quantum}. In our experiment, the flux pulses are shaped as $\text{Amp}_{i}(t) = A_{i}[\cosh(\eta/2) - \cosh(\eta t/T))]/[\cosh(\eta/2)-1]$, as illustrated in Fig.~\ref{Fig1}(c).
Here, $\eta = 10$ describes the steepness of the pulse edge, $T = 48$~ns is the CZ gate duration, and $A_{i}$ ($i = 1, 2$ and $c$) represents the pulse amplitudes of the qubits and couplers to be calibrated. Also, a total buffer of 16~ns is inserted on both sides of the CZ gate.

Figures \ref{Fig1}(d)-(f) describe the workflow for calibrating the CZ gate. The evolved circuits, including the SEA, the ELEA, and the CAFE, are shown in Appendix C. In the SEA, $Q_a$ is initialized to a superposition state with an $X/2$ gate. Then $2n$ CZ gates are applied to amplify gate errors, with $X$ gate and $Y$ gate on on $Q_a$  and $Q_b$, respectively, to cancel the accumulated single-qubit phases. The final $X/2$ gate on $Q_a$ projects the qubit onto the computational basis, enabling the readout of the accumulated conditional phase on $|11\rangle$. The ELEA circuit is inspired by the phase-averaged leakage error amplification (PALEA) circuit \cite{marxer2025above}. The circuit begins with an $X$ gate on both qubits to prepare state $|11\rangle$, followed by $n$ repetition units for leakage error amplification. Each unit consists of two CZ gates, with $X$ gates applied to both qubits before the first CZ gate, and an $X$ gate on $Q_a$ together with an $X_{12}$ gate on $Q_b$ before the second CZ gate.  Here, the $X_{12}$ gate represents the $\pi$ pulse acting on the subspace spanned by $|1\rangle$ and $|2\rangle$. In CAFE circuit, the qubits are initialized to a set of random states $|\psi\rangle$, followed by repetition of CZ gates to amplify the gate errors. Finally, the qubits are mapped back to the state $|00\rangle$ with an inverse circuit.
% By fitting the population of state $|00\rangle$ as a function of CZ gate count, the errors can be extracted.
% In calibrations, we repurpose this circuit to estimate the leakage error as a function of flux pulse amplitude, with a fixed CZ gate number.

We employ qubits $Q_a$ and $Q_b$ as an example to introduce the closed-loop calibration strategy. We set $A_c$ to a modest value and measure the population of $Q_a$ while scanning $A_{1}$ and $A_{2}$.
The result is presented in Fig.~\ref{Fig1}(d), with the dark area indicating an accumulated conditional phase on state $|11\rangle$.
We first select $A_1=0$, $A_2 = -0.099$ (marked in white) as an initial guess.
Next, in the intermediate calibration, we use the SEA and the ELEA circuits to scan $A_c$ and $A_2$ in alternation.
The results are shown in Fig.~\ref {Fig1}(e).
As the CZ gate number increases, $A_c$ and $A_2$ converge. In the fine calibration, we characterize the three-state readout of the two qubits, as shown in Fig.~\ref{Fig1}(f).
We replace the calibration of $A_c$ by measuring the population of $|02\rangle$ with the CAFE.
The results for applying 60 CZ gates in CAFE are shown in Fig.~\ref{Fig1}(g).
After the flux pulses are determined, the single-qubit phases $\phi_1$ and $\phi_2$ are compensated by adding virtual $Z$ gates on each qubit.

Notably, realizing CZ gates on large-scale processors faces the challenge of frequency crowding \cite{zhao2022quantum, 2021caiimpact}.
Frequency collisions between the interaction frequency and those of neighboring qubits or TLS defects lead to degradation of CZ gate fidelity.
Thus, an evaluation step is included in the calibration workflow.
When the fidelity of the CZ gate falls below a threshold $\mathcal{F}_{\text{th,CZ}}$, we update the choice of $A_1$ and $A_2$, then repeat the intermediate calibration and fine calibration steps.

\section{Benchmarking of CZ gates}

After the calibration, we perform standard randomized benchmarking (SRB) and interleaved randomized benchmarking (IRB) \cite{knill2008randomized, Magesan2012efficient} to determine the CZ gate fidelity of $Q_{a}$ and $Q_{b}$.
The population of state $|00\rangle$ decays exponentially with the two-qubit Clifford number, as shown in Fig.~\ref{Fig2}(a).
By fitting the decay rates, we obtain an error per Clifford (EPC) of $(0.228\pm0.004)\%$, with an average of $1.5$ CZ gates and $5$ $U_3$ single-qubit gates. The interleaved CZ gate fidelity is $(99.92\pm0.01)\%$.

We then measure the CZ gate fidelity with respect to the gate duration, as shown in the inset of Fig.~\ref{Fig2}(a). For gate duration ranging from $30$ to $48$~ns, comparable error rates below $0.1\%$ are obtained. However, further reduction in the gate duration leads to an increase in the error rate. Specifically, shorter CZ gate durations requires the coupler frequency to be tuned closer to the qubit frequencies, which induces non-negligible qubit population leakage to the coupler.
Meanwhile, residual short-time distortion of flux pulses also introduces coherent errors. This result indicates that there is a trade-off between incoherent errors and coherent errors.

% The population of states $|00\rangle$ versus the CZ number $m=2n$ can be fitted with
% \begin{align}
%     P_{m} &= \frac{1}{4} - \epsilon_{\text{SPAM}} \\
%     &+ \frac{(1-p)^{m}}{20}\bigg[16\prod\cos^2(m\theta_i)+16\prod\sin^2(m\theta_i)-1\bigg].\nonumber
% \end{align}
% Here, $\epsilon_{\text{SPAM}}$ represents the state preparation and measurement (SPAM) errors.
% $p$ represents the depolarizing rate,
% $\Delta\theta_i~(i=1,2,3)$  represent the phase errors of the imperfect CZ gate 

% \begin{equation}
%     \tilde{U}_{\text{CZ}} = \begin{pmatrix}
%         1&0&0&0\\0&e^{2i(\theta_2+\theta_3)}&0&0\\ 0&0&e^{2i(\theta_1+\theta_3)}&0\\
%         0&0&0&-e^{2i(\theta_1+\theta_2)}
%     \end{pmatrix}.
% \end{equation}

% \begin{equation}
%     \tilde{U}_{\text{CZ}} = \begin{pmatrix}
%         1&0&0&0\\0&e^{2mi(\theta_2+\theta_3)}&0&0\\ 0&0&e^{2mi(\theta_1+\theta_3)}&0\\
%         0&0&0&e^{2mi(\theta_1+\theta_2)}
%     \end{pmatrix}.
% \end{equation}

To analyze the gate errors, we apply the CAFE circuit (see Appendix C3 for details), as shown in Fig.~\ref{Fig2}(b), the fitted decoherent error and coherent error are $0.079\%$ and $0.007\%$, respectively.
The estimated CZ gate fidelity is $99.914\%$, in agreement with the IRB fidelity.
Note that, in addition to incoherent errors, the decoherent error also includes contributions arising from leakage to non-computational subspaces, primarily state $|02\rangle$, as shown in the inset of Fig.~\ref{Fig2}(b).
By fitting the population of state $|02\rangle$, we estimate the leakage error rate as $\epsilon_{\text{leak}}<0.001\%$. 

We also use CAFE with dynamical decoupling (DCAFE) by applying $X$ gates to both qubits after every pair of CZ gates, see inset of Fig. \ref{Fig2}(b). In this circuit, the incoherent error caused by low-frequency noise can be echoed out, suppressing the incoherent error of the CZ gate to $0.064\%$.
Meanwhile, the coherent error is reduced to $0.002\%$, resulting from the mitigation of single-qubit phase errors.

% Theoretically, Decreasing the CZ gate length enables a reduced incoherent error, contributes to a lower overall error rate. To this end, we vary the CZ gate length and measure the corresponding gate errors. The results are shown in Fig.~\ref{Fig3}(c). When the gate length is within $30-48$~ns, comparable error rates below $0.1\%$ are obtained. With further reduction of gate length, the error rate increases conversely. It indicates that there is a trade-off between incoherent errors and coherent errors. Reducing the CZ gate time necessitates higher coupler frequencies, which induces non-negligible qubit population leakage to the  coupler. In addition, the uncompensated short-time distortion of flux pulses also induce more coherent errors.

Furthermore, we extend this workflow to the parallel calibration and benchmarking of CZ gates. The spatial distribution of these errors across qubits is visualized in Fig.~\ref{Fig3}(a), revealing a correlation between CZ gate error rates and single-qubit gate error rates, both of which can be traced to short dephasing times (see Appendix B for details). The cumulative distribution function (CDF) of 65 single-qubit gate errors and 72 CZ gate errors are shown in Fig.~\ref{Fig3}(b), the median single-qubit gate error and CZ gate error are $0.05\%$ and $0.75\%$, respectively. The single-qubit error rate $\epsilon_{\text{SQ}}$ exhibits an exponential dependence on the inverse dephasing time, i.e., $\epsilon_{SQ}\propto\exp(-1/{T_2^{\text{XY8}}})$, with $T_2^{\text{XY8}}$ denoting the dephasing time measured with the $\text{XY8}$ dynamical decoupling sequence, as shown in Fig.~\ref{Fig3}(c). 

In analogy to the analysis of the $Q_{a}$-$Q_{b}$ CZ gate, we categorize errors for all CZ gates and show the resulting CDF in Fig.~\ref{Fig3}(d). 
The green area indicates the coherent error, exhibiting a median rate of $0.002\%$ that stems from phase errors in SU(4) decomposition.
The yellow area indicates the incoherent error induced by leakage, with a median rate of $0.016\%$. Another source of incoherent error, originates from energy relaxation and dephasing, is represented by the red area, with median rate of $0.5\%$. The overall error is consistent with the IRB error.

\section{Automated calibration}

\begin{figure}[t]
    \centering
    \includegraphics[]{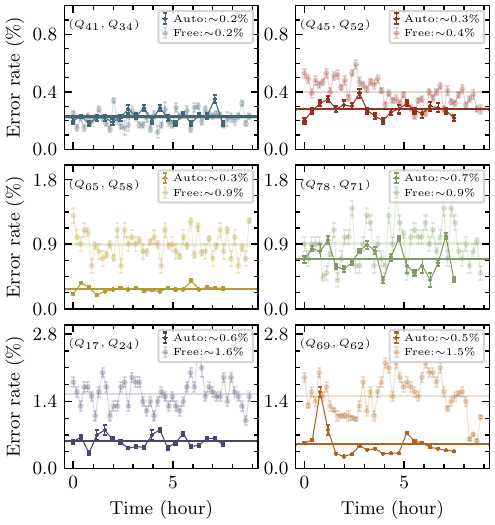}
    \caption{The error rates of six CZ gates in automated calibration (dark lines) and calibration-free (light lines) scenarios, each scenario consists of a 9-hour monitoring. The average error rates are plotted with horizontal lines. Error bars indicate the standard errors of the IRB fit.}   
    \label{Fig4}
\end{figure}

In scalable superconducting processors, qubits and couplers are typically biased away from their sweet-point frequencies, rendering the gates susceptible to environmental drifts.
Thus, automated calibration strategies are essential for maintaining high performance.
We show that, via the fine calibration step in the workflow, CZ gates with lower error rates and enhanced stability can be obtained.
We perform an 18-hour monitoring experiment on six qubit pairs and track their CZ gate error rates over 9 hours under automated calibration and calibration-free scenarios.
The automated calibration is performed every 0.5 hours.
The results are shown in Fig.~\ref{Fig4}, with dark and light lines representing automated calibration and calibration-free scenarios, respectively.
The horizontal lines represent the average error rates during the monitoring period.
Evidently, the average error rates decrease with automated calibration.
In particular, for the qubits that suffer from more fluctuations, such as $Q_{69}$ and $Q_{62}$, the automated calibration provides more benefit.

\section{Discussion and Outlook}

In summary, we have demonstrated a workflow for calibrating CZ gates on a scalable quantum processor. 
The high performance of this workflow benefits from the ELEA and CAFE circuits, facilitating precise quantification of leakage to the state $|02\rangle$.
With this workflow, we achieve a maximum gate fidelity of $(99.92\pm0.01)\%$ and a median fidelity of $99.25\%$ among 72 CZ gates.
In addition, the calibration can be executed automatically to enhance gate stability, making it compatible with the artificial intelligence at the current stage and offering the prospect of realizing fully automated calibration. 

To advance toward even higher fidelities, faster CZ gates are required to reduce incoherent errors. 
While challenging, this can be realized by optimizing the qubit interaction frequencies, CZ gate durations, and flux pulse shapes \cite{sun2021realization}, ensuring that the qubit and coupler undergo a full oscillation.
Our investigation of calibration workflow, along with fabrication technologies, paves the way for realizing fault-tolerant quantum computation with superconducting quantum processors.

\section{Acknowledgments}

Pei Liu gratefully acknowledges Dr. Yirong Jin for helpful discussions.
We acknowledge the support from the National Natural Science Foundation of China (Grants No.~92365206, No.~12404560, No.~12040561), and the Innovation Program for Quantum Science and Technology (No.~2021ZD0301802, No.~2024ZD0803301).

% \section{Data availability}

% The data generated in this study have been deposited in the zenodo database \cite{zenodo}.

%\bibliography{CZfidelity}


\clearpage
\onecolumngrid
\setcounter{figure}{0}
\setcounter{table}{0}
\setcounter{equation}{0}
\renewcommand{\thefigure}{S\arabic{figure}}
\renewcommand{\thetable}{S\arabic{table}}
\renewcommand{\theequation}{S\arabic{equation}}

% \title{Supplementary Information: High Precision Calibration workflow Achieves Above 99.9\% Fidelity on a Scalable Superconducting Processor}
% \maketitle

\section*{Appendix A: Experimental Setup}

\begin{figure}[h]
    \centering
    \includegraphics[]{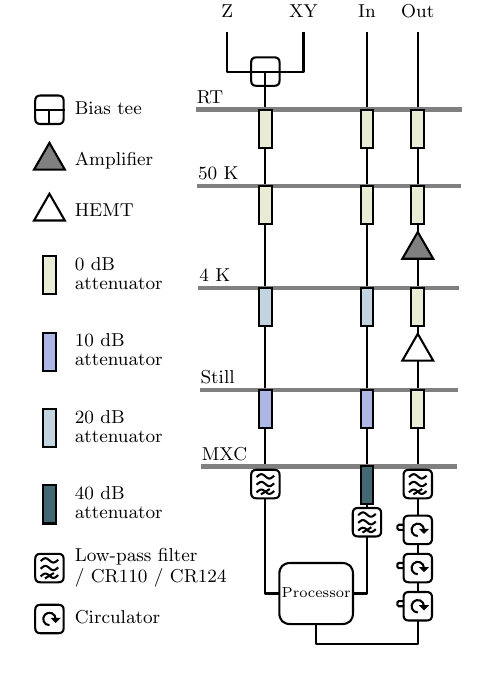}
    \caption{The schematic of the measurement system includes both the cryogenic and room-temperature setups.
    Three types of signals are shown: the control signals for the transmon qubit or coupler, the probe-in (In) signal, and the probe-out (Out) signal.
    For qubit control, the microwave drive (XY) and flux (Z) signals are combined using a bias tee, whereas for coupler control, only the Z signal is needed.
    Each control signal is filtered by a CR124 filter at the mixing chamber (MXC).
    For the probe-in signal, a $10$~GHz low-pass filter and an infrared filter (CR110) are mounted at the MXC, while the probe-out signal passes through another $10$~GHz low-pass filter and CR110.}
    \label{fig:setup}
\end{figure}

The experiment is performed on the quantum processor mounted in a ZL-DR~400 dilution refrigerator (DR) from \href{http://zlcryogenic.com}{ZL Cryogenic}, with a base temperature around 12.9~mK.
A total of $224$ high-density cables from \href{http://www.hermercs.com/index.asp}{Hermercs} are equipped in the DR, of which 196 cables are connected to the quantum processor.
The wiring configurations are shown in Fig.~\ref{fig:setup}.
Both the probe-in (In) and qubit drive (XY) signals are directly digital-synthesized by \href{http://zw-instrument.com}{ZWDX}~AWG-2000, and the flux pulses (Z) are generated by \href{https://www.chiphop.cn}{Chiphop}~ChipQ-AWG.
For qubit control, each XY signal is filtered with a band-pass filter, then combined with Z signal at room temperature via a bias-tee.
The signals are attenuated and filtered at multiple cold stages of the DR before transmitting into the quantum processor.
The probe-out (Out) signals are sequentially amplified by high electron mobility transistors (HEMTs) from ZWDX mounted at 4~K stage and a 50~K-amplifier from Ying-Ying, then modulated by analog-to-digital converters (ADC) from ZWDX.
All instruments and equipment utilized in this work are procured from Chinese suppliers.

\clearpage
\section*{Appendix B: Performance of the Superconducting Processor}

\begin{figure}[htbp]
    \centering
    \includegraphics[]{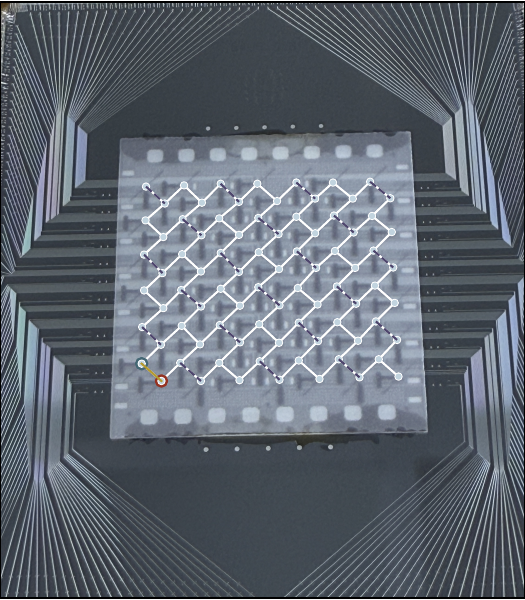}
    \caption{The optical image of the processor, with topology of qubits and couplers represented as a graph. The nodes (light blue dots) denote the transmon qubits and 
    edges (lines) represent interactions mediated by the transmon couplers. The couplers outlined by black dashed lines represent one of the six parallel‑manipulation groups. The CZ gate with fidelity above $99.9\%$ is performed on blue-circled qubit $Q_a$ and red-circled qubit $Q_b$, coupled via yellow-circled coupler $C$ on this processor.}
    \label{fig:chip}
\end{figure}

\begin{figure}[htbp]
    \centering
    \includegraphics[]{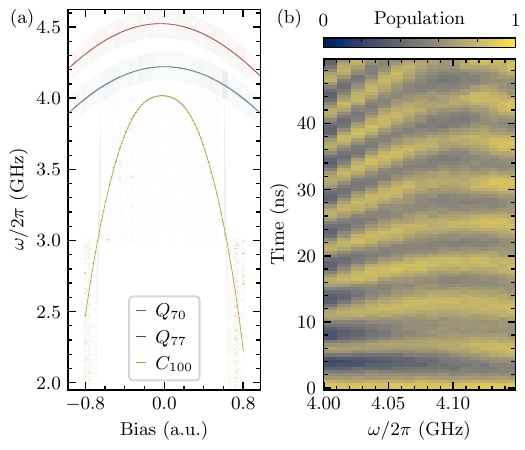}
    \caption{(a) The Spectrum of $Q_a~(Q_{70})$, $Q_b~(Q_{77})$, and $C~(C_{100})$. (b) The iSWAP interaction between $Q_{70}$ and $C_{100}$, with the interaction strength of $g_{qc}/2\pi\approx 87$~MHz.}
    \label{fig:spectrum}
\end{figure}

The experiment is carried out on the superconducting quantum processor \href{https://quafu-sqc.baqis.ac.cn/home}{$Shenglian$} consisting of 84 tunable transmon qubits \cite{baqisQuafuSuperconducting}.
An optical image of the processor is shown in Fig.~\ref{fig:chip}.
The qubits (light blue dots) are arranged in a hexagonal lattice, coupled via tunable transmon couplers (white lines). The couplers are arranged in six groups when calibrating and benchmarking CZ gates, one of the group is presented with black dashed lines. $Q_a$, $Q_b$ and $C$ in the main text are highlighted with blue, red and yellow, respectively.
Their frequencies versus bias are shown in Fig.~\ref{fig:spectrum}~(a).
The $Q_{a}$-$C$ interaction strength extracted from the iSWAP data is $g_{qc}/2\pi\approx 87$~MHz, as shown in Fig.~\ref{fig:spectrum}~(b).

\begin{figure*}[h]
    \centering
    \includegraphics[width=\linewidth]{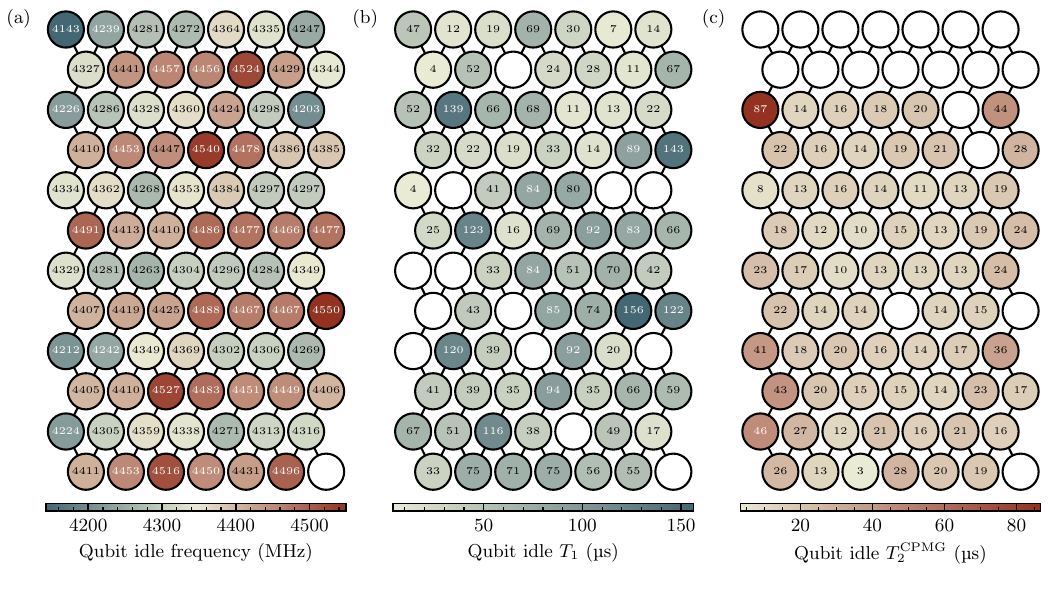}
    \caption{Heatmaps of (a) the qubit idle frequencies, (b) the energy relaxation time $T_1$, and (c) the decoherence time $T_2^{\rm CPMG}$.
    The blank circles indicate missing measurements or the fits outside the confidence interval.}
    \label{FigS2}
\end{figure*}

For the 65 qubits in the experiment, we consider various factors when designing the idle frequency arrangement strategy, including nearest-neighbor interactions, coherence properties, crosstalk, distortions, and so on. 
Fig.~\ref{FigS2}~(a) shows the idle frequencies for the 65 qubits. Their coherence properties, including $T_1$ and $T_2^{\text{CPMG}}$ are shown in Fig.~\ref{FigS2}~(b) and (c), the median $T_1$ and $T_2^{\text{CPMG}}$ are $50.8$~\textmu s and $16.9$~\textmu s, respectively.

\clearpage

\section*{Appendix C: Calibration circuits}

\begin{figure*}[ht]
    \centering
    \includegraphics[]{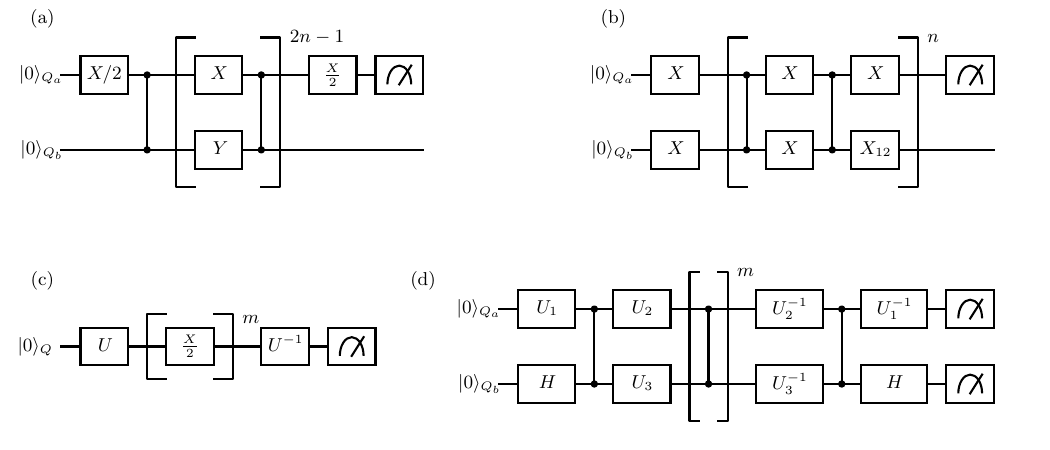}
    \caption{The quantum circuits applied in the calibration workflow.
    (a) The standard error amplification circuit for CZ gate calibration, which is sensitive to the $ZZ$ interaction and insensitive to single-qubit phase accumulation.
    (b) The echoed leakage error amplification circuit for CZ calibration, which avoids phase variations in $X_{12}$.
    (c) The context-aware fidelity estimation (CAFE) circuit for characterizing the single-qubit $X/2$ gate.
    (d) The CAFE circuit for characterizing CZ gate.}
    \label{FigS4}
\end{figure*}

\subsection*{1: Standard Error Amplification}
The standard error amplification (SEA) circuit is shown in Fig.~\ref{FigS4}~(a). $Q_a$ is initialized to a superposition state with an $X/2$ gate.
Then $2n$ repetitions of the CZ gate are applied to amplify the errors.
An $X$ gates on $Q_a$ and $Y$ gate on $Q_b$ are interleaved between adjacent CZ gates to cancel the effect of the accumulated single-qubit phase errors. 
The final $X/2$ gate on $Q_a$ projects the qubit onto the computational basis, enabling the readout of the accumulated conditional phase on $|11\rangle$.

\subsection*{2: Echoed Leakage Error Amplification}
The echoed leakage error amplification (ELEA) circuit is shown in Fig.~\ref{FigS4}~(b), which is inspired by the phase-averaged leakage error amplification (PALEA) circuit \cite{marxer2025above}.
In the PALEA, two $X$ gates prepare the qubits to state $|11\rangle$, after which $n$ repetitions of CZ gates followed by $X$ and $X_{12}(\phi)$ gates are applied.
Here, the $X_{12}(\phi)$ gate represents the $\pi$ pulse acting on the subspace spanned by $|1\rangle$ and $|2\rangle$, with relative phase $\phi$.
To estimate the leakage error, the PALEA requires random variation of $\phi$ in each repetition, and averaging over all the results.
In the ELEA, we add another CZ and two $X$ gates in the repetitive module, eliminating the requirement of varying the phase $\phi$, thus $X_{12} = X_{12}(0)$.
A detailed description of the ELEA and the PALEA is provided in the Appendix.

\subsection*{3: Context-Aware Fidelity Estimation}

The context-aware fidelity estimation (CAFE) circuit is proposed in Ref.~\cite{debroy2023context}, aiming to calculate the fidelity of a gate operation and analyze the error budget.
The CAFE circuit can be applied for characterizing both the single-qubit gate and the CZ gate.
Generally, in a CAFE circuit, the qubits are initially prepared to $|\psi\rangle$ belonging to an $k$-qubit 2-design $\{\psi_i\}$.
Then the target gate is repeated by $m$ times, and finally an inversion step is applied to prepare the qubits back to state $|0\rangle ^{\otimes k}$. The CAFE circuit for the single-qubit and two-qubit cases for characterizing the CZ gate is illustrated in Fig.~\ref{FigS4}~(c) and (d), respectively.

For the single-qubit case, the state $|\psi\rangle$ is realized by a single-qubit rotation $U$. For two-qubit case, the state $|\psi\rangle = (P_{0},
P_{1},P_{2},P_{3})^T$~($\sum_{i=0}^{3} |P_i|^2=1$ ) can be generalized by:

\begin{enumerate}[label=(\alph*)]
    \item Execute a $Y(\alpha)$ rotation to the first qubit $Q_1$ and a Hadamard gate to the second qubit. The rotation angle $\alpha = 2\arccos(S^{00}_{\psi})$, where $S_{\psi}$ is calculated by performing the singular value decomposition (SVD) of
    \begin{equation}
    M_{|\psi\rangle} = 
    \begin{pmatrix}
    P_0 & P_1 \\
    P_2 & P_3
    \end{pmatrix} =U_{\psi}S_{\psi}V_{\psi}^{\dagger}.
    \label{eq1}
    \end{equation}
    \item Perform a CZ gate to create an entangle state, noted as  $|E\rangle = (E_{0},E_{1},E_{2},E_{3})^T$ .
    \item Take SVD of 
    \begin{equation}
    M_{|E\rangle} = 
    \begin{pmatrix}
    E_0 & E_1\\
    E_2 & E_3
    \end{pmatrix} = 
    U_{|E\rangle}S_{|E\rangle}V^{\dagger}_{|E\rangle},
    \label{eq2}
    \end{equation}
     and apply single-qubit operations of $U_1=U_{|\psi\rangle}U_{|E\rangle}^{-1}$, $U_2=V_{|\psi\rangle}V_{|E\rangle}^{-1}$, respectively.
\end{enumerate}

With the conclusion in Ref \cite{debroy2023context}, when the gate operation $U$ depolarizes the qubits with probability $p$, the population of state $|0\rangle ^{\otimes k}$ and the gate number $m$ follows the relationship of 

\begin{equation}
    \mathcal{P}_m = \frac{1}{d}+(1-p)^m\frac{\big|\text{Tr}\big((U^{\dagger})^m\tilde{U}^{m}\big)\big|^2-1}{d(d+1)}-\epsilon_{\text{SPAM}},
    \label{eq3}
\end{equation}
here, $d = 2^k$ is the dimension of the Hilbert space, $\epsilon_{\text{SPAM}}$ represents the state preparation and measurement
(SPAM) errors. Without coherent errors, i.e., when $\tilde{U}=U$,
\begin{equation}
    \mathcal{P}_m = \frac{1}{d}+\frac{d-1}{d}(1-p)^m-\epsilon_{\text{SPAM}}.
    \label{eq4}
\end{equation}
We further derive the $\mathcal{P}_m$ when coherent error exists. 

For single-qubit gates,
\begin{equation}
    U(\theta,\phi) = \exp \left[-i\frac{\theta}{2}(\sigma_x\cos\phi+\sigma_y\sin\phi)\right].
    \label{eq5}
\end{equation}
Assume that the error originates from $\theta$,
\begin{equation}
{\rm Tr}\left([U^\dagger(\theta, \phi)]^mU^m(\theta+\Delta_\theta, \phi)\right)=2\cos\frac{m\Delta_\theta}2,
\label{eq6}
\end{equation}
then
\begin{equation}\label{sq_cafe}
{\mathcal P}_m=\frac12+\frac16(1-p)^m\left(4\cos^2\frac{m\Delta_\theta}2-1\right)-\epsilon_{\text{SPAM}}.
\end{equation}
It indicates that errors arising from the pulse duration or the amplitude.
Meanwhile, phase $\phi$ in the pulse originates from the digital waveform sampling and can therefore be treated as amplitude errors.
Frequency inaccuracies of the pulse can be regarded as imperfect rotation-angle errors,
i. e.,
\begin{equation}
    \begin{aligned}
        &\exp\left[-i\frac\theta2\left(\sigma_x\cos\phi+\sigma_y\sin\phi+\eta\sigma_z\right)\right]\\=&\cos\frac{\theta\sqrt{1+\eta^2}}2I-i\sin\frac{\theta\sqrt{1+\eta^2}}2\left(\sigma_x\cos\phi+\sigma_y\sin\phi+\eta\sigma_z\right)\frac1{\sqrt{1+\eta^2}},
    \end{aligned}
    \label{eq8}
\end{equation}
and can thus be analogized to the $\theta$-like errors.
Consequently, all single‑qubit gate errors can be expressed in Eq.(\ref{sq_cafe}).

For two-qubit case, when the imperfect CZ gate is expressed as 

\begin{align}
    &\tilde{U}_{\rm CZ}(\Delta_\alpha, \Delta_\beta, \Delta_\gamma) = \\\nonumber
&\begin{pmatrix}
\exp[-i(\Delta_\alpha+\Delta_\beta+\Delta_\gamma)]&0&0&0\\
0 & \exp[-i(\Delta_\alpha-\Delta_\beta-\Delta_\gamma)]&0&0\\
0&0& \exp[-i(-\Delta_\alpha+\Delta_\beta-\Delta_\gamma)]&0\\
0&0&0&-\exp[-i(-\Delta_\alpha-\Delta_\beta+\Delta_\gamma)]
\end{pmatrix},
\label{eq9}
\end{align}

we have

\begin{equation} \left|\text{Tr}\left((U^{\dagger})^m\tilde{U}^{m}\right)\right|^2= 16\left(\cos^2m\Delta_\gamma\cos^2m\Delta_\alpha\cos^2m\Delta_\beta+\sin^2m\Delta_\gamma\sin^2m\Delta_\alpha\sin^2m\Delta_\beta\right).
    \label{eq10}
\end{equation}

Apply Eq. (\ref{eq10}) to Eq. (\ref{eq3}), the population of states $|00\rangle$ versus the CZ number $m$ can be fitted with
\begin{equation}
    \mathcal{P}_{m} = \frac{1}{4} - \epsilon_{\text{SPAM}} 
    + \frac{(1-p)^{m}}{20}\bigg[16\left(\cos^2m\Delta_\gamma\cos^2m\Delta_\alpha\cos^2m\Delta_\beta+\sin^2m\Delta_\gamma\sin^2m\Delta_\alpha\sin^2m\Delta_\beta\right)-1\bigg].
    \label{eq11}
\end{equation}

The decoherent errors and coherent errors are calculated by 
\begin{align}
\epsilon_\text{decoh} &= 1-\frac{\mathcal{P}_1(\Delta_\alpha = \Delta_\beta= \Delta_\gamma = 0)}{1-\epsilon_{\text{SPAM}}},\nonumber\\ 
\epsilon_\text{coh}& = 1-\frac{\mathcal{P}_1(p=0)}{1-\epsilon_{\text{SPAM}}}.
\label{eq12}
\end{align}

\section*{Appendix D: Benchmarking CZ gates with different durations}
\begin{figure*}[h]
    \centering
    \includegraphics[width=\linewidth]{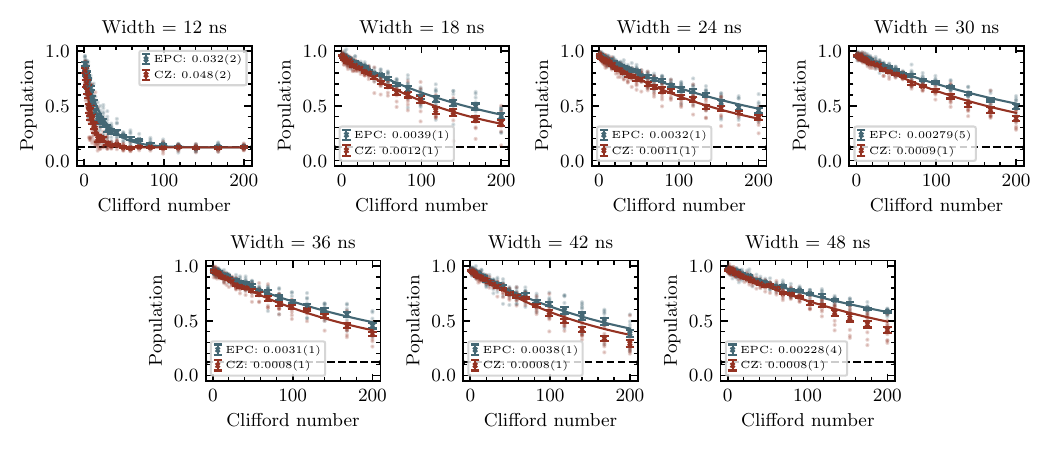}
    \caption{Randomized benchmarking results with different widths, from $12$~ns to $48$~ns in steps of $6$~ns.}
    \label{FigS3}
\end{figure*}

As mentioned in the main text, the errors in the CZ gate face a trade-off between the effects of decoherence and the leakage to the coupler. The standard and interleaved randomized benchmarking results shown in Fig.~\ref{FigS3} support this argument and also imply that long-range imperfections remain in the system.

\end{document}